\documentclass[aps,twocolumn,superscriptaddress,showpacs, pre]{revtex4-1}

\usepackage{graphicx,amsfonts,amsmath,color,amsbsy,amssymb}
\usepackage{epstopdf}

\newcommand{\eps}{\boldsymbol{\epsilon}}

\begin{document}
\title{Stresses in  curved nematic membranes}


\author{J.A. Santiago}\email{jsantiago@correo.cua.uam.mx}
\affiliation{Departamento de Matem\'aticas Aplicadas y Sistemas\\ 
Universidad Aut\'onoma Metropolitana Cuajimalpa\\
Vasco de Quiroga 4871, 05348 Cd. de  M\'exico, MEXICO}

\vspace{12pt}


\begin{abstract}

Ordering configurations of a director field on a curved membrane induce 
stress. In this work, we present a theoretical framework 
to calculate the stress tensor and the torque  as a consequence of the nematic ordering;
we use the variational principle and invariance of the energy under 
Euclidean motions. Euler-Lagrange equations of the membrane as well 
as the corresponding  boundary conditions also appear as  natural results.
The stress tensor found includes  attraction-repulsion forces
between defects;  likewise,  defects are attracted to patches with the same sign in  gaussian curvature.
These forces are mediated by the Green function of
Laplace-Beltrami operator of the surface. In addition, we find   non-isotropic forces 
that involve derivatives of the Green function and the gaussian curvature, even in the normal direction to the membrane.
We examine the case of  axial membranes to analyze the spherical one. For spherical vesicles 
we find the modified Young-Laplace law  as a consequence of the nematic texture. 
In the case of spherical cap with defect at the north pole, we find that the force 
is repulsive respect to the north pole, indicating that it is an unstable equilibrium~point.

\end{abstract}

\maketitle

\section{Introduction}
When extrinsic couplings  of Frank's energy describing liquid crystals on curved membranes are neglected, one finds that
defects interact with each other through the Green function of the Laplace-Beltrami operator of the surface\cite{vortices, kamien};
they also have interactions with the membrane itself and a bulk term appears describing the interaction of the gaussian curvature
of the membrane mediated by the Green  function.
Clearly, these interactions induce stresses along the membrane which in turn responds by modifying its shape:
the interest in determining the shape of biological membranes because
it is related to  specific  functions of the cell\cite{udo, kozlov}.
The distribution of stress along the membrane plays a relevant role, whether its shape can change or remain fixed. 
If the shape of the membrane is frozen, the amount of topological charge is determined precisely by the topology of the membrane 
through the Hopf-Poincar\'e and Gauss-Bonnet theorems\cite{spivak, yan}.
The nematic texture with  defects determines how the stress is distributed along the membrane.\\
The stress tensor has been calculated in several different ways:  in \cite{stress} and using a variational principle the authors 
find it in the case of fluid membranes;  in \cite{napoli2010} and using an elegant and general geometric formalism, the authors  find this tensor for very general schemes that can be applied to the relevant case of elastic membranes coated with nematic textures; in \cite{fournier} the author finds the stress tensor of the bending energy, 
examining deformations respect to a flat membrane. Remarkably, in \cite{jem} the  author finds 
this tensor in a novel way by using auxiliary variables, avoiding the tedious calculations of 
deforming the geometric objects involved.\\
The first main result of this article is the covariant  stress tensor of Frank's energy in the so-called limit of one constant, denoted
$\kappa_A$. Although extrinsic effects is a subject of great interest\cite{napoli, santangelo, Frank, vergo},
in  the model we examine,  the extrinsic couplings are not taken into account, 
but instead interactions between topological defects and the gaussian curvature of the surface 
are explicitly introduced. This model can be seen as the dominant approximation  of an effective 
energy that includes extrinsic corrections in curvature.\\
The stress tensor found exhibits the forces  in the nematic membrane:
Two like(unlike) charge defects repeal(attract)  each other. Defects are attracted to  patches with the same sign in 
gaussian curvature,
the interaction being through the Green function  $(- 1/\nabla^2)$   of the surface.   
We also find  non-isotropic forces that involve derivatives of the Green function and the gaussian curvature, 
a result that exhibits a more complex  non-isotropic forces  than those described above.\\
Using this theoretical framework,
we also find the  covariant Euler-Lagrange equation for  the nematic energy. 
This equation describes the shape of the membrane that is coupled with the 
configuration of the director field. It  is the covariant form of the von K\'arm\'an equation\cite{landau}, 
to which it is reduced when we use the Monge approach.
In the calculation of deformations of the nematic energy, we have found that the
tangential deformations  do not
imply only a boundary term, that is because  this energy is not invariant under reparametrizations: 
the presence of  the nematic texture implies elastic stresses tangent to the membrane.
Moreover, when the variational principle is implemented, the
boundary conditions for a free edge appear naturally.
We write these conditions in terms of  geometrical information of the edge curve.\\
As a relevant example, we obtain the stress tensor in the case of axially symmetric membranes. 
If the membrane is closed, we find the corresponding Young-Laplace law, eq.\eqref{UA}, which gives us the relationship with the pressure difference $P$ 
between inside and outside. Although this is also a relevant result of this work, this expression still 
depends on the nematic texture on the vesicle.\\
Therefore, we analyze the spherical case that has been studied not only from a theoretical point of view 
but also experimentally\cite{shin, vinel, lopez, fernandez}. 
Placing  $+1$ defect at  each pole of the spherical vesicle,   
we find the relationship of $P$ with the radius $R$ of the membrane, the surface  tension $\sigma $ and 
the nematic constant $\kappa_A$~to~be
\begin{equation}
P= \frac{2\sigma}{ R} \left(1- \frac{ \alpha\kappa_A}{2\sigma R^2 }  \right), 
\end{equation}
where $\alpha\in(0, 1)$ is  constant that depends of the nematic texture.
Notice the negative sign of the nematic  correction, 
unlike the positive sign by bending rigidity\cite{fournier2008}:
while the elastic force of the membrane adds to the surface tension,  the nematic force subtracts it.\\
Taking into account that elastic membranes have $ \sigma \sim 10^{-2} J/m^2 $ and for typical liquid crystals
$\kappa_A \sim 4.1 \times 10^{- 21} J $,  to have a nematic correction of at least $10\%$, the sphere must be
$R\sim 0.7 nm$. Nevertheless, some liquid crystals have  $\kappa_A \sim 10^{-11}N $ and     $\sigma\sim 10^{-5}J/m^2$, 
\cite{morris,  musevic}; for these liquid crystals we have a nematic correction of $50\%$ if $R\sim 1 \mu m$,  a reasonable size 
in which the nematic correction can be observed

If the membrane is not closed, the stress tensor is conserved on its surface. We must also 
take into account that in addition to the integrated gaussian curvature 
over the area, the edge curve determines the topological charge through the Gauss-Bonnet theorem.
We analyze two examples within the spherical cap, one of them  with charge $+1$ and the other one with charge $+1/2$ 
at the north pole. The result  we get is that the force on any horizontal loop  is repulsive respect to the
point defect at the pole.\\
The rest of the paper is organized as follows: in section \ref{PRE} we  give a brief review
to describe the Frank energy  on a curved surface, in the limit of one constant. In section \ref{SHAPENOETHER} 
we obtain the response of the energy to small deformations of the embedding function. 
To avoid confusion in the reading  we have separated the calculation of the normal and tangential deformation. 
In section \ref{BOUNCON} the boundary conditions are obtained. The key point here is to project the edge deformations along the Darboux basis.
By using the invariance of the energy under translations and rotations,  in section \ref{STTOR} 
we find the stress tensor and the torque. 
In section \ref{AXIALLY}, the case of membranes with axial symmetry is examined, and then some results for 
the spherical case are obtained.
We finished the article with a brief summary in section \ref{SUMM}. 
Most of the long calculations  have been written in several appendices at the end of the paper.
\section{Nematic energy}\label{PRE}
Let us consider  a surface in $\mathbb{R}^3$ of coordinates ${\bf x}= ( x^1, x^2, x^3 )$.  The surface
is parametrized by   $\xi^a$, through the embedding functions ${\bf x}= {\bf X} (\xi ^a)$. The induced metric on the surface 
is given by $g_{ab}={\bf e}_a \cdot {\bf e}_b$, the euclidean inner product in $\mathbb{R}^3$ of the tangent vectors ${\bf e}_a=\partial_a {\bf X}$ to the surface.
The unit normal vector to the surface is defined as
$
{\bf n}= {\bf e}_1\times {\bf e}_2 /\sqrt{g},
$
where $g={\rm det}\, g_{ab}$.
The covariant derivative compatible with the induced metric will be denoted  $\nabla_a$. 

Frank's energy describes the ordering of a unit
director field $\boldsymbol{\eta}$. This energy includes  the effect of splay, twist and bend the field along the surface. 
In the limit of one coupling constant the Frank energy can be written as~\cite{NelsonNano} 
\begin{equation}
F=\frac{ {\kappa}_A}{2} \int_{\cal M} dA (\nabla_a \eta^b)^2. \label{EFE}
\end{equation}
The integral involves
the infinitesimal area element on the patch $\cal M$ given by $dA= \sqrt{g}\, d^2\xi$, and the coupling with the extrinsic curvature 
has been neglected; nevertheless  by using theoretical and numerical simulations methods,
some recent works have taken  into account  extrinsic effects\cite{napoli, vergori, santangelo}.

A  convenient alternative  route to  describe this field theory,  is in terms of the spin connection ${\bf\Omega}= {\bf e}^a \Omega_a$, a vector  valued function 
defined in the tangent space of the surface \cite{david}, whose
fundamental property is its relationship with the gaussian curvature 
\begin{equation}
\nabla\times {\bf\Omega} =  {\cal R}_G {\bf n}.
\end{equation}

We define an  orthonormal basis $\eps_\alpha$, $\alpha=\{1, 2\}$, such that the field $\boldsymbol{\eta}$
can be   written in terms of  the angle $\Theta$ with $\eps_1$:
\begin{eqnarray}
\boldsymbol{\eta} &=& \eta^\alpha \eps_\alpha,\nonumber\\
&=&\cos\Theta \eps_1 +\sin\Theta\eps_2.
\end{eqnarray}
The spin connection is defined by $\eps_1\cdot \nabla_a \eps_2= \Omega_a$,  and with that  we have an alternative way of write
the nematic energy \eqref{EFE} as \cite{david}
\begin{equation}
F= \frac{\kappa_A}{2} \int_{\cal M} dA  g^{ab} (\partial_a \Theta-\Omega_a) (\partial_b\Theta-\Omega_b).\label{UL}
\end{equation}
Euler-Lagrange equation of the field $\nabla^a(\nabla_a\Theta  -\Omega_a)=0$, implies that
a scalar field $\chi$ exists such that $-\varepsilon^{ab} \nabla_b \chi= \nabla^a \Theta - \Omega^a$, where
$\varepsilon^{ab}= \epsilon^{ab}/ \sqrt{g}$.
The presence of topological defects screening by the gaussian curvature of the membrane is the source of
this field:  
\begin{equation}
 -\nabla^2 \chi = \rho_D(\xi)- {\cal R}_G, \label{EQE}
\end{equation}
where $\rho_D(\xi )= \sum_i q_i \delta( \xi-\xi_i)$ is the charge density.
A formal solution of \eqref{EQE} can be written as
\begin{equation}
{\chi}=\sum_i q_i  G(\xi, \xi^i ) -{\cal U},\label{CHI}
\end{equation}
where $G(\xi, \zeta)$ denotes the Green function associated with the Laplace-Beltrami operator  on the surface such that
\begin{equation}
-\nabla^2 G(\xi, \zeta)= \frac{\delta(\xi-\zeta)}{\sqrt{g}},
\end{equation}
and 
\begin{equation}
{\cal U}(\xi) =\int_{\cal M} dA_\zeta G(\xi, \zeta){\cal R}_G(\zeta),
\end{equation}
defines the geometric potential.
The  energy can thus  be written as
\begin{eqnarray}
F&=& \int_{\cal M} dA  (\nabla_a \chi)^2, \nonumber\\
&=& \int_{\cal M} dA\nabla_a (\chi \nabla^a \chi) + \int_{\cal M} \, dA \chi (- \nabla^2  )\chi.\label{OPP}
\end{eqnarray}
The first integral in \eqref{OPP} is a boundary term and the second one is the
bulk term that  can be developed as
\begin{eqnarray}
&&\int_{\cal M} dA\,  \chi \,  (-\nabla^2)\chi = \int_{\cal M} dA \,  \chi (\xi ) \, [ q_i \delta (\xi-\xi^i) - {\cal R}_G ], \nonumber\\
&&= q_i q_j G(\xi^i, \xi^j) + q_i\, {\cal U} (\xi^i) + \int_{\cal M} dA\,  {\cal U}(\xi) {\cal R}_G(\xi).  \label{ENR}
\end{eqnarray}
From this we see that defects interact with each other through the Green function,  we also see that
the geometric potential plays the role of an external electric field.
The last term is the interaction energy between the gaussian curvature mediated by the Green function.  \\
In the next section, the shape equation and boundary conditions of  the  functional energy  
\begin{equation}
{\cal H} =\sigma\int_{\cal M} dA  + \kappa_A \int_{\cal M} dA \,  \chi\, (- \nabla^2) \chi   + \sigma_b\oint_{\cal C} ds, 
\label{ENT}
\end{equation}
will be obtained,  $\sigma$ is the surface tension of the membrane patch ${\cal M}$ and 
$\sigma_b$ the linear tension of its boundary~${\cal C}$.

\section{Shape equations and Noether charges}\label{SHAPENOETHER}
To find the shape equation,  we obtain  the response of  the energy  (\ref{ENT}),  to small 
deformations of the embedding functions, ${\bf X}\rightarrow {\bf X} + \delta{\bf X}$.  
We project the deformation  into its tangential and normal  to the surface 
\begin{eqnarray}
\delta{\bf X}&=& \delta_{\parallel} {\bf X} +  \delta_\perp {\bf X}, \nonumber\\
&=&\Phi^a {\bf e}_a + \Phi\,  {\bf n}.
\end{eqnarray}
As a first step, we get  from  eq.\eqref{EQE}:
$-\delta \nabla^2 {\chi} =\delta\rho_D- \delta {\cal R}_G $.\\
Now, when the   area of the surface is modified, the total defects  can also be modified. Nevertheless,  
if the total area remains fixed, local deformations of the surface implies deformations
of the charge density without further changes in the total defects.
Thus, since the total charge $Q=\int_{\cal M} dA\,  \rho_D$ is preserved, we have that
$\delta Q=\int_{\cal M} (\delta dA) \rho_D + \int_{\cal M} dA\, \delta\rho_D=0$, in such a way that locally 
\begin{equation}
\delta \rho_D= -\rho_D (\nabla_a\Phi^a +K \Phi ),  \label{DEFDEN}
\end{equation}
where we used the area deformation,  $\delta dA= dA (\nabla_a\Phi^a + K\Phi)$.

Let us first get  the normal variation of the nematic energy.
This  deformation can be obtained by using  the commutator 
$[\delta_\perp , \nabla^2 ]{\chi} =J_\perp$ where $J_\perp=- 2K^{ab}\Phi \nabla_a \nabla_b {\chi}+ \nabla_b[ ( Kg^{ab} -2K^{ab})\Phi] \nabla_a  {\chi}$, see
\cite{deformations},  so that we can 
write $-\nabla^2 \delta_\perp {\chi} = \delta_\perp\rho_D -    \delta_\perp {\cal R}_G + J_\perp$ and 
deformation of the  energy gets
\begin{equation}
\delta_\perp F= -\int_{\cal M} dA \, K \, ( \rho_D + {\cal R}_G) \, {\chi} \,   \Phi + 
\int_{\cal M} dA\, [ J_\perp  -2\delta_\perp {\cal R}_G   ] {\chi}, 
\end{equation}
where we used  the normal deformation of the charge density, according to eq.\eqref{DEFDEN}:
$ \delta_\perp \rho_D= - \Phi K \rho_D$.
Deformation of the gaussian curvature has also been calculated  as \cite{deformations}
\begin{equation}
\delta_\perp  {\cal R}_G = - {\cal R}_G K\Phi + (K^{ab}-g^{ab}K ) \nabla_a \nabla_b \Phi \label{normaldef}.
\end{equation}
After some algebra  and several integrations by parts we have
\begin{equation}
\delta_\perp F= \int_{\cal M} dA \, {\cal E}_\perp\, \Phi + \int_{\cal M} dA \nabla_a Q_\perp^a, 
\end{equation}
where the Euler-Lagrange derivative of the nematic energy and the Noether charge $Q^a_\perp$ are given by
\begin{eqnarray}
&&{\cal E}_\perp = 2(Kg^{ab} - K^{ab} )\nabla_a\nabla_b {\chi} +    (2K^{ab}-  K g^{ab} )\nabla_a {\chi} \nabla_b {\chi },     \nonumber\\
&&Q^a_\perp = -2(K^{ab}- Kg^{ab}) {\chi}\nabla_b\Phi  \nonumber\\
&&+ [(Kg^{ab}- 2K^{ab}  ) {\chi} \nabla_b {\chi} + 2(K^{ab}- g^{ab} K )  \nabla_b {\chi}] \Phi. \label{Qa}
\end{eqnarray}
This expression for the Noether charge has not been completed;
tangential deformation is needed and  as we shall see, it is not just a boundary term.\\
Let's now get the tangential deformation.
For the  scalar curvature we have (see appendix)
\begin{equation}
\delta_{\parallel}{\cal R} =\Phi^a\nabla_a {\cal R}.\label{R}
\end{equation}
Notice  that the tangential deformation $\delta_\parallel F$ is not only a boundary term, 
this happens  because the nematic energy is not reparameterization invariant.
The presence of the director field breaks out this property of the bending energy.
To prove this, we see that the commutator with the laplacian is given by
$
[\delta_\parallel, \nabla^2 ]{\chi} = J_\parallel, 
$
where now, 
\begin{equation}
J_\parallel=( -\nabla^2 \Phi^a + {\cal R}_G \Phi^a )\nabla_a {\chi} - 2 (\nabla^a\Phi^b )\nabla_a\nabla_b {\chi}.\label{CCM}
\end{equation}
By using this commutator we have that
$
-\nabla^2 \delta_\parallel {\chi}= J_\parallel+\delta_\parallel \rho_D - \delta_\parallel {\cal R}_G,
$
and thus the tangential deformation does depend on the Green function. 
By using that $\delta_\parallel \rho_D= -\rho_D \nabla_a\Phi^a$ and
proceeding as in the case of the normal deformation we have
\begin{equation}
\delta_\parallel F = \int_{\cal M} dA\,  ( {\cal E}_ a\Phi^a + \nabla_a Q^a_\parallel), \label{deft}
\end{equation}
where we have identified
\begin{eqnarray}
{\cal E}_a &=& 2( \rho_D + {\cal R}_G \chi) \nabla_a\chi,  \nonumber\\
Q^a_\parallel&=& \Phi^b [ \nabla^a( \chi\nabla_b \chi ) -2\chi \nabla^a\nabla_b \chi - \delta^a_b (\rho_D + {\cal R}_G) \chi        ]
 \nonumber \\
&& - \chi \nabla_b\chi  \nabla^a \Phi^b. \label{QP}
 \end{eqnarray}
In order to obtain the Euler-Lagrange equation of the energy \eqref{ENT}, we write its bulk deformation
\begin{equation}
\delta {\cal H}= \int_{\cal M} dA\,  \boldsymbol{\cal E} \cdot \delta{\bf X} + \int_{\cal M} dA \nabla_a Q^a,
\end{equation}
where the Euler-Lagrange derivative 
\begin{equation}
\boldsymbol{\cal E}= (\kappa_A {\cal E}_\perp + \sigma K) {\bf n} + {\cal E}_a {\bf e}^a, 
\end{equation}
and the Noether charges  in $Q^a= \kappa_A Q^a_\perp +(\kappa_A Q^a_\parallel + \sigma \Phi^a)$, 
are given by eqs.\eqref{Qa} and \eqref{QP}. In equilibrium we have  $\boldsymbol{\cal E}=0$, and therefore its components  must vanish: ${\cal E}_\perp+ \sigma K =0={\cal E}_a$.

An interesting fact occurs if there are no defects on the membrane; in such a case  we have that $\chi= - {\cal U}$ and 
${\cal E}_a=0$ implies that $ \nabla_a {\cal U}=0$, so that the Euler-Lagrange  equation simplifies to
\begin{equation}
 K(\sigma  + 2\kappa_A  {\cal R}_G)=0, 
\end{equation}
and therefore,  minimal surfaces or hyperbolic-like surfaces 
are solutions to the Euler-Lagrange equation \cite{Frank,  Giomi, Gil}.  Notice  that  this result has been obtained 
by  deforming the energy functional ${\cal H}$, eq.\eqref{ENT}, which contains the function $\chi$. If instead of doing that, 
one deforms \eqref{UL}, which involves  $\Omega_a$, we get an apparently different result\cite{indios}. We will tackle
this interesting point in a future work. \\
As we will see below, from  the Noether charge $Q^a$ we can find both, the  stress tensor
and the torque; these  can be found when writing
explicitly a translation and  rotation of the embedding function.  
Before that, let us find the boundary conditions that appear naturally
in the variational principle.
\section{Boundary conditions}\label{BOUNCON}
According to the previous section, in equilibrium shapes,
deformation of energy  \eqref{ENT} including the boundary terms, is given by
\begin{equation}
\delta{\cal H} =  \kappa_A \oint_{\cal C} ds\,   l_a Q^a +  \sigma \oint_{\cal C} ds \, l_a \Phi^a +  \sigma_b\,  \delta \oint_{\cal C}  ds,   \label{DEFF}
\end{equation}
and thereby the boundary conditions will be obtained by doing $\delta{\cal H}=0$.\\
The calculation involves the Darboux basis adapted to the boundary ${\cal C}$ parametrized by arc length $s$ \cite{docarmo}.
Deformation of the boundary  can be projected as
\begin{eqnarray}
\delta{ \bf X} &=&\Phi^a {\bf e}_a + \Phi {\bf n}, \nonumber\\
&=& \phi {\bf T}+ \psi {\bf l} + \Phi {\bf n},
\end{eqnarray}
where we have defined the scalar funcions $\Phi^a T_a = \phi$ and $\Phi^a l_a=\psi$.
Therefore, deformation of the unit tangent can be written as
\begin{eqnarray}
\delta {\bf T} &=&\dot\phi {\bf T} + \dot{\psi} {\bf l} + \dot\Phi {\bf n}
+ \phi \dot{\bf T} + \psi \dot{\bf l} + \Phi \dot{\bf n}, \nonumber\\
&=& ( \dot\phi - \kappa_g\psi -\kappa_n \Phi   ) {\bf T} + ( \dot\psi +\kappa_g \phi + \tau_g \Phi    ){\bf l} \nonumber\\
&+& (\dot\Phi + 	\kappa_n\phi - \tau_g \psi ) {\bf n}.
\end{eqnarray}
where $\kappa_g$ is the geodesic curvature, $\kappa_n$ the normal curvature, and $\tau_g$ the geodesic torsion
of the bondary, see App.\eqref{DARFRA}. The point means derivative respect to arclength.  
Then we obtain \cite{santiago}
\begin{eqnarray}
\delta\oint_{\cal C}\, ds &=& \oint_{\cal C}\, ds\,  {\bf T} \cdot\delta {\bf T }, \nonumber\\
&=& \oint_{\cal C} ds\, ( \dot\phi - \kappa_g\psi -\kappa_n \Phi   ), \nonumber\\
&=& \Delta\phi- \oint_{\cal C} ds\, (  \kappa_g\psi + \kappa_n \Phi   ).
\end{eqnarray}
where $\Delta\phi= 0$ for a closed curve.  Thus, $\delta L $
does not include deformation along the unit tangential  vector.
According to  \eqref{Qa} and \eqref{QP}
we have $ l_a Q^a =l_a(Q_\perp^a + Q_\parallel^a  )$. If we write
\begin{eqnarray}
Q_\perp ^a&=& M^{ab} \nabla_b\Phi + M^a \Phi\nonumber\\
Q_\parallel^a&=& N^a{}_b\Phi^b + N_b\nabla^a\Phi^b,
\end{eqnarray}
where 
\begin{eqnarray}
M^{ab}&=&2 ( Kg^{ab}- K^{ab} ) \chi \nonumber\\
M^a&=&[( Kg^{ab} - K^{ab} ) (\chi -2) - K^{ab}\chi  ] \nabla_b \chi, \nonumber\\
N^{ab}&=&  \nabla^a\chi\nabla^b\chi - \chi \nabla^a\nabla^b\chi - g^{ab}(\rho_D+ {\cal R}_G )\chi, \nonumber\\
N^a&=& -\chi \nabla^a \chi. 
\end{eqnarray}
we have the boundary conditions, see Appendix\eqref{DARFRA}
\begin{eqnarray}
-\kappa_A \frac{d}{ds}\left(  l_a M^{ab} T_b  \right) +\kappa_A l_a M^a -\sigma_b \kappa_n&=&0, \nonumber\\
\kappa_A(l_a N^{ab}l_b   + N^b\nabla_l l_b  ) +\sigma-\sigma_b\kappa_g&=&0, \nonumber\\
l_a M^{ab}l_b&=&0, \nonumber\\
N^bl_b&=&0, \nonumber\\
N^b T_b&=&0.
\end{eqnarray}
where we have used that on the boundary
$
\nabla_a\Phi =l_a \nabla_l \Phi + T_a \dot\Phi, 
$
and 
$
\nabla_b\Phi^a=T_b \dot\Phi^a + l_b\nabla_l \Phi^a, 
$
and the fact  that on the boundary, the independent  deformations are given by the scalar functions $ \psi , \phi, \Phi $.

\section{Stress and torque}\label{STTOR}
How the stress is distributed along a membrane is the information that is encoded
in the stress tensor\cite{stress, deserno}. To find it, we write the deformation of the energy as
\begin{equation}
\delta {\cal H}= \int_{\cal M} dA\,  \boldsymbol{\cal E} \cdot \delta{\bf X} + \int_{\cal M} dA \nabla_a Q^a,
\end{equation}
where the Euler-Lagrange derivative 
$  \boldsymbol{\cal E}= ({\cal E}_\perp + \sigma K) {\bf n} + {\cal E}_a {\bf e}^a $
and the Noether charges $Q^a= Q^a_\perp +Q^a_\parallel$, 
are given by eqs.\eqref{Qa} and \eqref{QP}. In equilibrium we have that   $\boldsymbol{\cal E}=0$, that implies   ${\cal E}_\perp=0={\cal E}_a$.

If the energy is invariant under reparametrizations, then its
tangential deformation is a  boundary term and  $ {\cal E}_a $ vanish identically; 
however, if the energy does not have this invariance, as in the case of the nematic energy, 
these terms are not trivial as we  see in eq.(\ref{deft}).

On the other hand, invariance of energy under translations implies that
$\delta H = 0$, so that  locally we have
\begin{equation}
 \boldsymbol{\cal E} = \nabla_a {\bf f}^a \label{SSRR}
\end{equation}
where ${\bf f}^a$ is the stress tensor.
In equilibrium, the conservation law of the stress $\nabla_a{\bf f}^a=0$ is fulfilled and 
thus ${\bf F} = \oint_{\cal C}\,  ds \, {\bf f}^a l_a$, is  a conserved vector field along the surface; it is identified as
the force acting on the curve ${\cal C}$  parametrized by arc lenght $s$ with normal $l_a$.
The tangential derivatives  ${\cal E}_b$ will be relevant when coupled with 
crystalline order through the strain deformation\cite{landau, cristal}.\\
In the case of a membrane that encloses a certain volume $V$, we must add the term $PV$ to the energy, where $P$ 
is the pressure difference  between the interior and the exterior. In that case the stress tensor is not conserved but 
$ \nabla_a {\bf f}^a= P\, {\bf n}$,  in such a way that
\begin{equation}
\oint_{\cal C} ds\, {\bf f}^al_a = \int_{\cal M} dA P\, {\bf n}. \label{YLP}
\end{equation}

\subsection{Stress}
\begin{figure}
\includegraphics[scale=0.30]{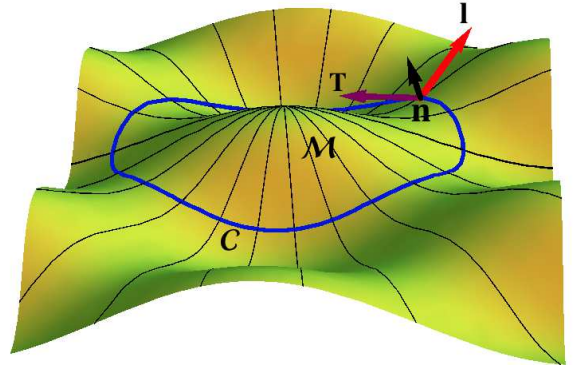}
\caption{A $+1$ defect on the top of a mountain. At any point of the curve ${\cal C}$, the force per unit length along ${\bf l}$ 
is given by $-(\sigma + \kappa_A\chi{\cal R}_G )$. In addition, there is the anisotropic force
$\kappa_A l^al^b(\chi \nabla_a\nabla_b \chi - \nabla_a\chi \nabla_b\chi)$ along {\bf l}.  Darboux frame adapted  to the curve is shown: ${\bf T}$  the unit tangent, 
the unit normal to the surface ${\bf n}$,   and ${\bf l}={\bf T} \times {\bf n}$.}
\label{DFC}
\end{figure}

Under an infinitesimal translation $\delta {\bf X}= {\bf a}$, we have that $ \Phi= {\bf a}\cdot {\bf n}$, and  $\Phi^a= {\bf a}\cdot {\bf e}^a$; 
we also see that $\nabla_b \Phi= {\bf a}\cdot K_b{}^c {\bf e}_c$. Substituting in eqs.(\ref{Qa}) and (\ref{QP}),   we find the stress tensor as
\begin{equation}
{\bf f}^a= (f^{ab}_\perp + f^{ab}_\parallel) {\bf e}_b + (f^a_\perp +f^a_\parallel ) {\bf n}, \label{STRESS}
\end{equation}
where the coefficients are given by
\begin{eqnarray}
&&f^{ab}_\perp= - g^{ab}(\sigma  + 2\, {\chi}\, {\cal R}_ G),   \nonumber\\
&&f^{ab}_\parallel= {\chi}\nabla^a\nabla^b{\chi} - \nabla^a {\chi}\nabla^b{\chi}  + g^{ab}( \rho_D+ {\cal R}_G){\chi}, \nonumber\\
&&f^b_\perp= - (Kg^{ab} - 2K^{ab} ) {\chi}\nabla_a {\chi} -  2 (  K^{ab} -g^{ab}K)     \nabla_a {\chi},\nonumber\\
&&f^b_\parallel=- K^{ab}{\chi} \nabla_a{\chi}. \label{FLL}
\end{eqnarray}
We have verified that the relationship (\ref{SSRR})  with the Euler-Lagrange derivatives  is fulfilled, this guarantees that both, the expression for the
stress tensor  and the shape equation  are self-consistent.\\
Let  ${\bf x}(s)= {\bf  X} ({\xi}^a(s))$, be a curve $\cal C$  parametrized  by arc length on the surface, see Fig.\eqref{DFC}; as before, we identify the Darboux basis adapted to it:
${\bf T}= T^a{\bf e}_a$ its tangent vector  and ${\bf l}= l^a{\bf e}_a$ the outward  pointing unit vector,   such that
${\bf l}= {\bf T}\times {\bf n}$.
The force per unit of length can be written as
\begin{equation}
{\bf f}^al_a=   F_T {\bf T} + F_l {\bf l}+ F_n {\bf n},\label{FORCE}
\end{equation}
where $F_T= l_a T_b f^{ab}$, 
$F_l= l_al_b f^{ab} $, and $F_n = l_af^a $. We get
\begin{eqnarray}
F_l&=& -f 
+ l_al_b ( {\chi} \nabla^a \nabla^ b{\chi} - \nabla^a {\chi} \nabla^b {\chi} ), \nonumber\\
F_T&=&l_aT_b ( {\chi} \nabla^a \nabla^ b{\chi} - \nabla^a {\chi} \nabla^b {\chi} ), \nonumber\\
F_n&=&l_b(K^{ab}-g^{ab}K )( \chi - 2 ) \nabla_a\chi. \label{FLTN}
\end{eqnarray}
Note that $ F_l $ includes $f =\sigma - (\rho_D - {\cal R}_G) \chi $.
This force can be written explicitly 
\begin{eqnarray}
&&-f = - \sigma  + \sum_{i\neq j}q_iq_j \delta({\bf x}-{\bf x}^i)G({\bf x}, {\bf x}^j) \nonumber\\
&-& \sum_iq_i G({\bf x}, {\bf x}^i){\cal R}_G - \sum_iq_i \delta ({\bf x}-{\bf x}^i) {\cal U} + {\cal U}{\cal R}_G . \label{SIGMA}
\end{eqnarray}
The second term is  the force on the charge $ q_i $ due to $ q_j $, it is given by
$ q_iq_j G ({\bf x}^i, {\bf x}^j)$, this force is repulsive(attractive)  between  defects  with  like(unlike) charge.
Similarly, the third term is the force  on the point ${\bf x}$ (of gaussian curvature  ${\cal R}_G$), caused by the presence of $q_i $ at the point ${\bf x}^i $:
defects are attracted  to points with the same sign of gaussian curvature. These interactions are mediated by the Green function.
The fourth term  is a self-force at the point $ {\bf x}^i$ with the gaussian curvature at the same point.

The total force along ${\bf l}$ includes the anisotropic  stress  $\kappa_A l_al_b(  {\chi} \nabla^a \nabla^ b{\chi} - \nabla^a {\chi} \nabla^b {\chi} )$, along $\bf l$ and $ {\bf T}$. 
Finally, there is  also a force $F_n$ along the unit normal to the surface as given in eq.\eqref{FLTN}.  None of these forces has been reported so far.

\subsection{Torque}
Taking now an infinitesimal rotation $\delta {\bf X}= {\bf b}\times {\bf X}$, we have that  $\Phi= {\bf b}\cdot {\bf X}\times {\bf n}$ 
and $\Phi_a= {\bf b}\cdot {\bf X}\times {\bf e}_a$. Therefore, we can write
\begin{eqnarray}
\nabla^b\Phi &=& {\bf b}\cdot \left( {\bf e}^b\times {\bf n} + K^{ab}\, {\bf X}\times {\bf e}_a \right).\nonumber\\
&=&{\bf b} \cdot \left(  \varepsilon^{ab} {\bf e}_a +  K^{ab} {\bf X}\times {\bf e}_a \right).
\end{eqnarray}
Similarly we have
\begin{equation}
\nabla_b\Phi_a={\bf b}\cdot (\varepsilon_{ba} {\bf n} -K_{ab}\, {\bf X}\times {\bf n}).
\end{equation}
where now $\varepsilon_{ab}=\sqrt{g} \epsilon_{ab}$.
Deformation of the energy  under a rotation is then given by  \cite{stress}
\begin{equation}
\delta{\cal H}=\int_{\cal M} dA\,  \boldsymbol{\cal E}\cdot ({\bf b}\times {\bf X} ) + \int_{\cal M} dA \nabla_a {\bf m}^a, 
\end{equation}
where 
\begin{equation}
{\bf m}^b= {\bf X} \times {\bf f}^b + {\bf s}^b, \label{TORQUE}
\end{equation}
being ${\bf f}^b$  the stress tensor \eqref{STRESS},    and 
\begin{equation}
{\bf s}^b= 2 (K_a{}^b - \delta^b_a K )\, \chi\, \varepsilon^{ac} {\bf e}_c  +\varepsilon^{ab} \chi \nabla_a \chi \, {\bf n}.
\end{equation}
In equilibrium we have $\boldsymbol{\cal E}=0$ so that  ${\bf m}^a$ is conserved as a
consequence of invariance under rotations. The first term in eq.\eqref{TORQUE} is the orbital 
torque while ${\bf s}^{b}$ can be seen as an intrinsic torque.
If we use the fact that  $\varepsilon^{ac}{\bf e}_c= l^a {\bf T} - T^a {\bf l} $, then we obtain the 
intrinsic torque in the Darboux basis  along a curve on the membrane.

\section{Axial nematic membranes}\label{AXIALLY}
Let us see the case of axial surfaces parametrized 
as
\begin{eqnarray}
{\bf X}(l, \phi ) &=&( \rho(l)\cos\phi, \rho(l)\sin\phi,  h(l)),\nonumber\\
&=& \rho \boldsymbol{\rho} + h{\bf k}
\end{eqnarray}
where $\boldsymbol{\rho}=(\cos\phi, \sin\phi, 0) $ is a unit  radial vector field,
and ${\bf k}=(0, 0, 1)$.
The tangent vectors to the surface can be found to be
\begin{eqnarray}
{\bf e}_l &=& (\rho'\cos\phi, \rho'\sin\phi, h'),  \nonumber\\
&=& \rho' \boldsymbol{\rho} +h' {\bf k}. \nonumber\\
{\bf e}_\phi &=& ( -\rho\sin\phi, \rho\cos\phi, 0), \nonumber \\
&=& \rho\, \boldsymbol{\phi}.
\end{eqnarray}
where $\boldsymbol{\phi}=(-\sin\phi,  \cos\phi, 0)$ is the unit azimuthal vector and $'$ denotes derivative respect to $l$.
The induced metric on the surface can be written as
\begin{equation}
g_{ab}d\xi^ad\xi^b = dl^2 +  \rho^2 d\phi^2, \label{DSS}
\end{equation}
where we have taken  the parameter  $l$ along the meridians to be the arc length  such that $h'^2+\rho'^2=1$.
The unit normal  to the surface ${\bf n}=\boldsymbol{\phi}\times {\bf e}_l$,  is given by
\begin{eqnarray}
{\bf n}&=&  (h' \cos\phi, h' \sin\phi, -\rho' ),\nonumber\\
&=&  h' \boldsymbol{\rho} - \rho'  \,{\bf k}.
\end{eqnarray}
The second fundamental form  can be written as
\begin{equation}
K_{ab} d\xi^ad\xi^b= -\frac{\rho'' }{h'}dl^2 + \rho h' d\phi^2
\end{equation}
whereas the mean curvature $K=h'/\rho - \rho''/h'$ and 
the gaussian curvature, 
$
{\cal R}_G=- \rho'' / \rho.
$
Let $\eps_1= \boldsymbol{\phi}$ and $\eps_2=\rho' \boldsymbol{\rho} + h' {\bf k}$ be the unit basis so that
the components of the spin connection are given by $\Omega_l= 0$ and $\Omega_\phi=\rho'$.
Along a horizontal curve we have $l_l={\bf l}\cdot {\bf e}_l=1 $ and $T_l=0$ so that
in these coordinates the coefficients (\ref{FORCE}) of the force $l_l{\bf f}^l $ 
per unit length on a horizontal loop can also  be written as
\begin{equation}
l_l{\bf f}^l= ( F_l \rho' + F_n h' ) \boldsymbol{\rho} + ( F_l h' - F_n \rho' ) {\bf k}, \label{FG}
\end{equation}
where we have
\begin{eqnarray}
F_l &=& -\sigma+ \left( \rho_D+  \frac{\rho''}{\rho} \right)\chi + [ \chi \chi'' - (\chi')^2 ],  \nonumber\\
F_n &=&  \left(- \frac{h'}{\rho} \right)( \chi - 2 ) \chi'   ,\nonumber\\
F_T&=& 0.
\end{eqnarray}
We note that although $\rho= \rho( l)$  by the axial symmetry, 
in a general setting,  the presence of the nematic texture implies that the coefficients depend on both variables $(l, \phi)$  on the surface, 
through the function $\chi$. This force has radial and vertical components. The total vertical force on the loop is then
\begin{eqnarray}
{\bf F} (l) &=& {\bf k} \int_0^{2\pi} d\, \phi \, \rho ( F_l h' - F_n \rho' ), \nonumber\\
 &=& {\bf k} [  h'   \langle F_l \rangle  -  \rho' \langle F_n \rangle       ] , 
 \end{eqnarray}
 where we have denoted    $\langle F \rangle =\int_0^{2\pi} d\phi \rho F $.  If the membrane
is a closed surface we must  take into account the pressure difference $P$ between the inside and outside
to the nematic membrane.
The equation \eqref{YLP} is then 
\begin{equation}
2\rho ( h'  F_l -  \rho' F_n )=- P\rho^2 , \label{UA} 
\end{equation}
where we have taken $\rho(0)=0$. This equation must be satisfied
for each value of $l$ in the domain considered;  it is the corresponding Young-Laplace law.

\subsection{Spherical particles}
\begin{figure}[htbp]
\begin{center}
\begin{minipage}[t]{8cm}
\centering
\includegraphics[scale=0.27]{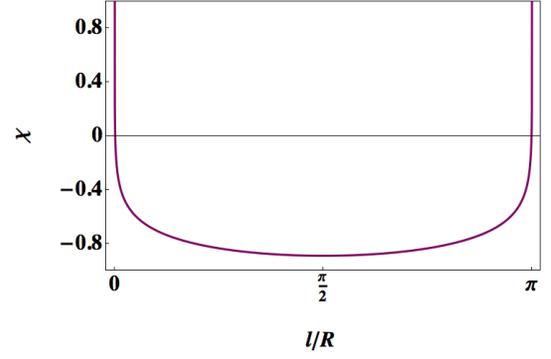} 
\end{minipage}
\caption{The function $\chi$ in eq.\eqref{XL} with   $+1$ defect at each  pole  where singularities appear.}
\label{CHI-SPHE}
\end{center} 
\end{figure}

Without nematic texture in the membrane such that $F_l=-\sigma$ and $F_n=0$,  eq.(\ref{UA})  reduces to
$ 2\rho h' \sigma = P\rho^2$. 
By using that $ h'=\sqrt{ 1-\rho'^2 }$,  and taking the simplest case such that $P$ is a constant we obtain
\begin{equation}
\rho (l) =\frac{2\sigma}{P} \sin\left( \frac{Pl}{2\sigma}  \right),
\end{equation}
which is the representation of a sphere with radius $R= 2\sigma / P$, this is
corresponding Young-Laplace equation, which relates the surface tension $\sigma$, the pressure $P$ and the 
radius of the sphere $R$.
Let us find the  corresponding law  in the presence of the nematic texture.  From eq.\eqref{UA},  we see that it is necessary 
to calculate the function $\chi$ that involves the Green function on the sphere.
To this, write the metric in isothermal coordinates
\begin{equation}
ds^2= \omega (dr^2 + r^2 d\phi^2),  
\end{equation}
where $r>0 $,  $\phi \in [0, 2\pi]$, 
and $\omega$  the conformal factor\cite{willmore}.
Comparison with the induced metric in axial coordinates \eqref{DSS}  gives
 \begin{equation}
dl^2=\omega dr^2, \,\, \,\,\, \, \omega r^2= \rho^2.
\end{equation}
That is,  $\log r= \int dl/\rho +C$.
Let $\xi=(l, \phi ) $ and $ \zeta=(\ell, \varphi ) $ and write  the Green function that satisfies the equation
\begin{equation}
-\Big[ \frac{1}{\rho} \partial_l \left(  \rho\,  \partial_l  \right) + \frac{1}{\rho^2 }\partial_\phi^2 \Big] G(\xi, \zeta) = \frac{1}{\rho}\delta( l-\ell )\delta(\phi- \varphi ), 
\end{equation}
replacing with isothermal coordinates 
 $u= (r, \phi)$, 
gets into
\begin{equation}
-\nabla^2 G =-\frac{1}{\omega} \nabla_u^2 G(u, u') = \frac{1}{\omega r}\delta(r-r')\delta(\phi-\phi'). \label{GP}
\end{equation}
The last equality in eq.\eqref{GP}   implies  the Green function in isothermal coordinates 
\begin{equation}
G(  u, u' )= -\frac{1}{4\pi}\log [ r(l)^2 +r(\ell) ^2 -2r(l) r(\ell) \cos(\phi -\varphi) ]. \label{GFF}
\end{equation}
If the surface is closed, the singularities that appear into  the Green function can be eliminated
if we subtract both
$\bar G(\xi)= (1/A)\int dA_\zeta G(\xi, \zeta) $
and   $\bar G(\zeta) $.
Let us look explicitly the example of the sphere;  parametrize it as
\begin{eqnarray}
\rho(l) &=& R\sin(l/R),\nonumber\\
h(l)&=&-R\cos(l/R). 
\end{eqnarray}
where $l\in [0, \pi R]$.
If we choose  $r( \pi R/2 )= R$ then we have
\begin{equation}
r(l)= R\tan \left( \frac{l}{2R} \right), \label{RS}
\end{equation}
and we can obtain
\begin{eqnarray}
\bar G(\xi)&=&\frac{1}{A}\int_0^{\pi R} d\ell\,  \rho (\ell) \int_0^{2\pi} d\phi \,\, G(\xi, \zeta), \nonumber\\
&=&- \frac{1}{8\pi R^2} \int_0^{\pi R} d\ell\,  \rho(\ell) \,  \log r_>^2, \nonumber\\
&=&\frac{1}{4\pi} \log \cos^2 \left(\frac{l}{2R}\right), \label{GAB}
\end{eqnarray}
where $r_>$  refers  to the larger value between  $r(l)$ and $r(\ell)$.
The Green function can then  be written as
\begin{eqnarray}
G( \xi, \zeta)&=&-\frac{1}{4\pi} \log [  \sin^2 (l/2R) \cos^2( \ell /2R) \nonumber\\
&&+ \sin^2 (\ell /2R) \cos^2 (l/2R) \nonumber\\
&&- \frac{1}{2} \sin (l/R) \sin (\ell /R) \cos(\phi -\varphi ) ]. \label{green}
\end{eqnarray}
Therefore, as shown in appendix \eqref{GREEN},  the geometric potential is simply given by
$
{\cal U}=1.
$
Thus, with a charge $+1$  at each pole, the function $\chi $ can be written as
\begin{equation}
\chi (l) =-\frac{1}{4\pi} \log\left[ \sin^2 \left( \frac{l}{2R}\right)  \cos^2 \left(\frac{l}{2R} \right)\right]-1. \label{XL}
\end{equation}
Notice that as a  consequence of topological defects at the poles,  singularities  in  eq.\eqref{XL} appear, see Fig.\eqref{CHI-SPHE}

Now, since eq.\eqref{UA} is fulfilled for $l\in (0+\epsilon, \pi R - \epsilon )$, 
where $ \epsilon$  is related with the core of defects, it can be rewritten  as
\begin{equation}
P=\frac{2\sigma}{R }\left( 1- \frac{ \alpha }{R^2}  \frac{\kappa_A}{2 \sigma }  \right),\label{YP}
\end{equation}
where $\alpha$ is a fixed number $ \in (0, 1)$,  that is obtained from
\begin{equation}
\alpha= \dot\chi ( \chi -2  ) - \chi + ( \chi \ddot\chi - \dot\chi^2   ),\label{ALPHA}
\end{equation}
where the dot means derivative respect to $x=l/R$.
We see that the surface tension has been modified by the presence of the nematic texture with $+1$ defects at the poles. 
As mentioned in the introduction, for spherical membranes with  $R \sim 1\mu m$, coated with some  
liquid crystals,  the nematic correction is
about $50 \%$.

\begin{figure}[htbp]
\begin{center}
\begin{minipage}[t]{4cm}
\centering
\includegraphics[scale=0.35]{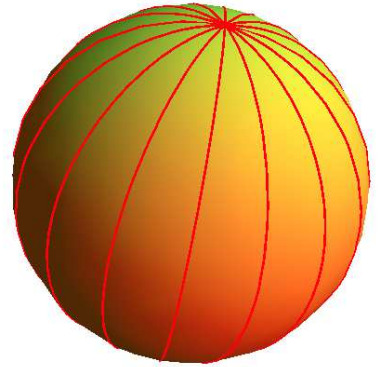} 
\end{minipage}
\caption{A spherical {\it particle} with radious $R$ and  $+1$ defects on opposite sides; with this nematic texture, the relationship between the parameters
is given by the Young-Laplace law~\eqref{YP}. }
\label{CT}
\end{center} 
\end{figure}

\begin{figure*}[htbp]
\begin{center}
\begin{minipage}[t]{8cm}
\centering
\includegraphics[scale=0.28]{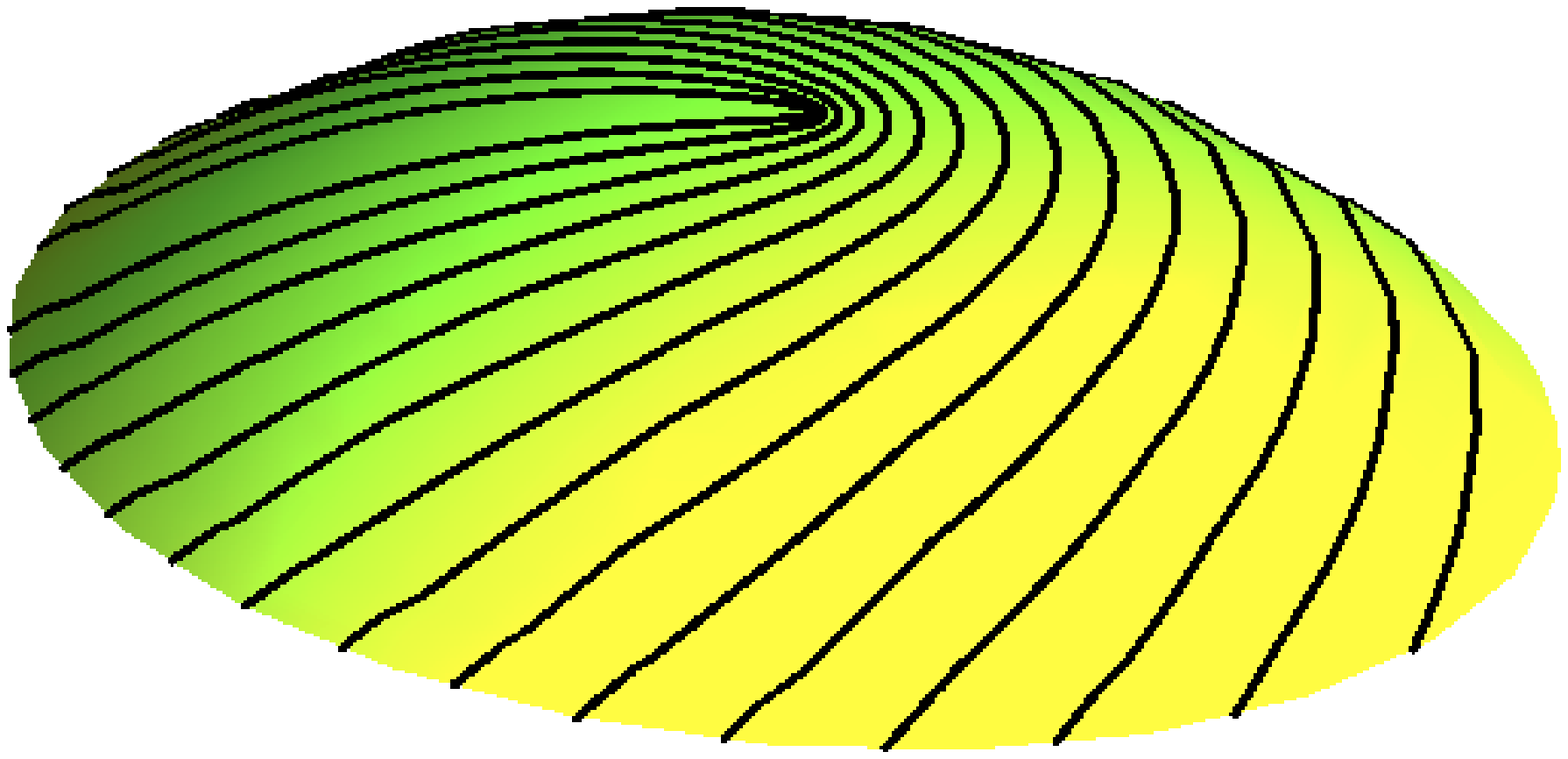} 
\end{minipage}
\begin{minipage}[t]{8cm}
\centering
\includegraphics[scale=0.28]{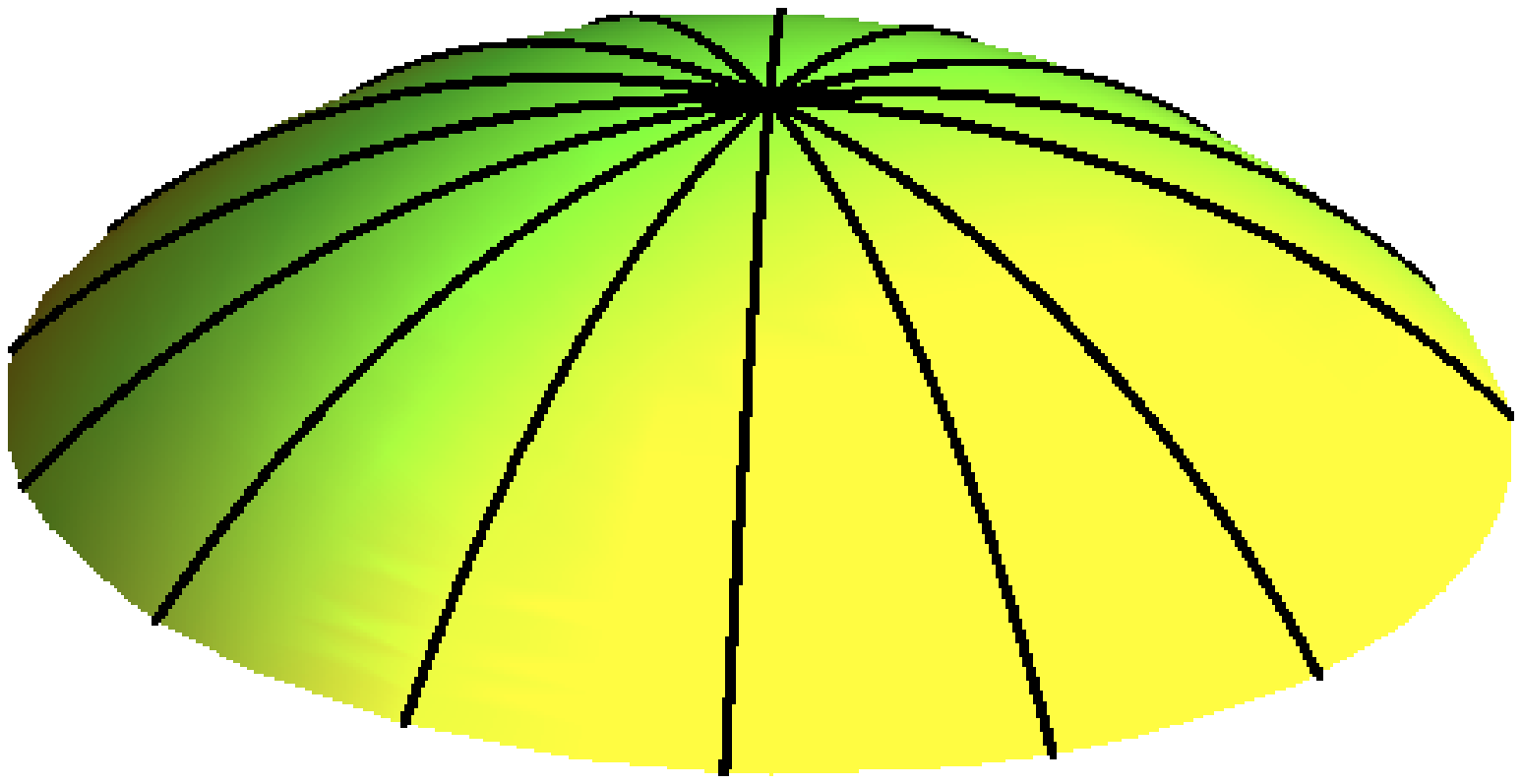}
\end{minipage}
\caption{Nematic texture on spherical sheets with boundary at $l_0=2R\arcsin(\sqrt{3/2}/2 )$  and  $l_0= \pi R/2$ respectively.
Gauss-Bonnet theorem implies  defects with $q=+1/2$ and $q=+1$ on them.  }
\label{SPP}
\end{center} 
\end{figure*}

In the case of a spherical cap  ${\cal M}$, Gauss-Bonnet implies that
\begin{equation}
\int_{\cal M} dA\, {\cal R}_G + \oint_{\cal C}\kappa_g ds= 2\pi Q, 
\end{equation}
where $\kappa_g $ is the gaussian curvature of the boundary curve ${\cal  C} $ parametrized by arc length $s$. The sum of these 
integrals is equivalent to the charge $Q$ of  defects into the surface. Integration of the gaussian curvature gives
\begin{eqnarray}
\int_{\cal M} dA {\cal R}_G&=&\frac{2\pi}{R^2} \int_0^{l_0} dl R \sin(l/R),  \nonumber\\
&=&4\pi \sin^2 (l_0/2R).
\end{eqnarray}
If the boundary is the parallel  $l=l_0 $, then we find
\begin{equation}
\oint_{\cal C} \kappa_g ds= 2\pi \cos(l_0/R),
\end{equation}
and therefore, the total charge on the spherical cap is given by
\begin{equation}
Q= 4\sin^2 (l_0/2R) - 1.\label{GB}
\end{equation}
For a half sphere  $l_0=\pi R/2$, we have $Q=1$, in such a case, the boundary is a geodesic curve with $\kappa_g=0$; 
a cap with  $l_0=\pi R/3 $ as boundary point,  has a nematic texture with $Q=0$.  
If $l_0= 2\pi R/3$, then we have $Q=2$. Notice that $Q=1/2$ if $l_0=2R\arcsin(\sqrt{3/2}/2 )\sim  5\pi R/12$ and there is not  $l_0$ such that  $Q=-1$.
Two of these caps with their  nematic texture are shown in fig.\eqref{SPP}
\\
For each of these spherical shells the  
Green function  is given by eq.\eqref{GFF} while $r (l)$ by eq.\eqref{RS}, but now we  must  to impose boundary  conditions on the 
Green function at  $l=l_0$. Under Dirichlet boundary conditions it reads
\begin{equation}
G(\xi, \zeta ) =-\frac{1}{4\pi} \log \left[ \frac{ r^2(l) +r(\ell)^2 -2r(l) r(\ell) \cos (\phi-\varphi)   }{  \frac{r(l)^2r(\ell)^2}{r_0^2}+r_0^2 -2r_0r(\ell) \cos (\phi-\varphi)  }         \right],
\end{equation}
where $r_0=r(l_0)$.
After making some integrations we can find  the geometric potential $ {\cal U}$ as
\begin{eqnarray}
{\cal U} &=&- \log\left[   \frac{  \cos^2 (l/2R) \sin^2 (l/2R) } {  \sin^2(l_0/2R)     }     \right] \nonumber\\
&&+\cos(l/R) \log\left[ \frac{ \tan^2(l/2R) }{ \tan^2 (l_0/2R)} \right].
\end{eqnarray}
For a half spherical cap, $r_0=R$ and $ \sin (l_0/2R)=1/ \sqrt{2}$, and thus we get
\begin{eqnarray}
{\cal U} &=&- \log\left[ 2   \cos^2 (l/2R) \sin^2 (l/2R)   \right] \nonumber\\
&&+ \cos(l/R) \log\left[  \tan^2 (l/2R) \right].
\end{eqnarray}
If the boundary is at the  point $l_0=2R\arcsin ( \sqrt{3/2}/2 ) $, we obtain the geometric potential as
\begin{eqnarray}
{\cal U} &=&- \log\left[ \frac{8}{3}   \cos^2 (l/2R) \sin^2 (l/2R)   \right] \nonumber\\
&&+ \cos(l/R) \log\left[ \frac{5}{3} \tan^2 (l/2R) \right].
\end{eqnarray}
Fig.\eqref{CT} shows these  geometric potentials:  in order to minimize the energy, 
defects must to be at $l=0$; nevertheless as we shall see, it is an
unstable equilibrium point.
For a half sphere such that $l_0=\pi R/2$  and doing $\ell=0$,  (defect at the pole $\zeta=\zeta_N$) we have
\begin{equation}
G( \xi, \zeta_N )=-\frac{1}{4\pi } \log [ \tan^2 (l/2R)].
\end{equation}
If the  boundary is at $l_0/R= 2\arcsin(\sqrt{3/2}/2)$ and defect at the north pole, we obtain
the Green function as
\begin{equation}
G( \xi, \xi_0 )=-\frac{1}{4\pi } \log \left[\frac{5}{3} \tan^2 (l/2R) \right].
\end{equation}

Since the membrane is not closed, then $ {\bf F} $ in eq.\eqref{FG}, 
is a conserved quantity, in particular  we evaluate it at the equator
of the half sphere. In this case  $F_l$ can be written as
\begin{equation}
F_l=-\sigma -\frac{1}{R^2} ( \chi -\chi\ddot\chi +\dot\chi^2 ), 
\end{equation}
where the dot means  derivative respect to $ x=l/R$. 
The force on a horizontal loop is thereby given by 
 \begin{eqnarray}
&&{\bf F}= \langle F_l \rangle {\bf k}, \nonumber\\
&&= -2\pi R\sigma \left( 1 + \frac{\kappa_A}{ \sigma R^2} \frac{C}{2}  \right) {\bf k}, \label{FHH}
\end{eqnarray}
where $C\sim 0.72$ for  half sphere with $q=1$ at the north pole,
and $C\sim 0.49 $ for  spherical cap with defect $q=+1/2$ at the pole.
This force acts to elongate the shape of the membrane towards cylindrical forms\cite{mackintosh, nguyen, powers}.

\begin{figure*}[htbp]
\begin{center}
\begin{minipage}[t]{8cm}
\centering
\includegraphics[scale=0.28]{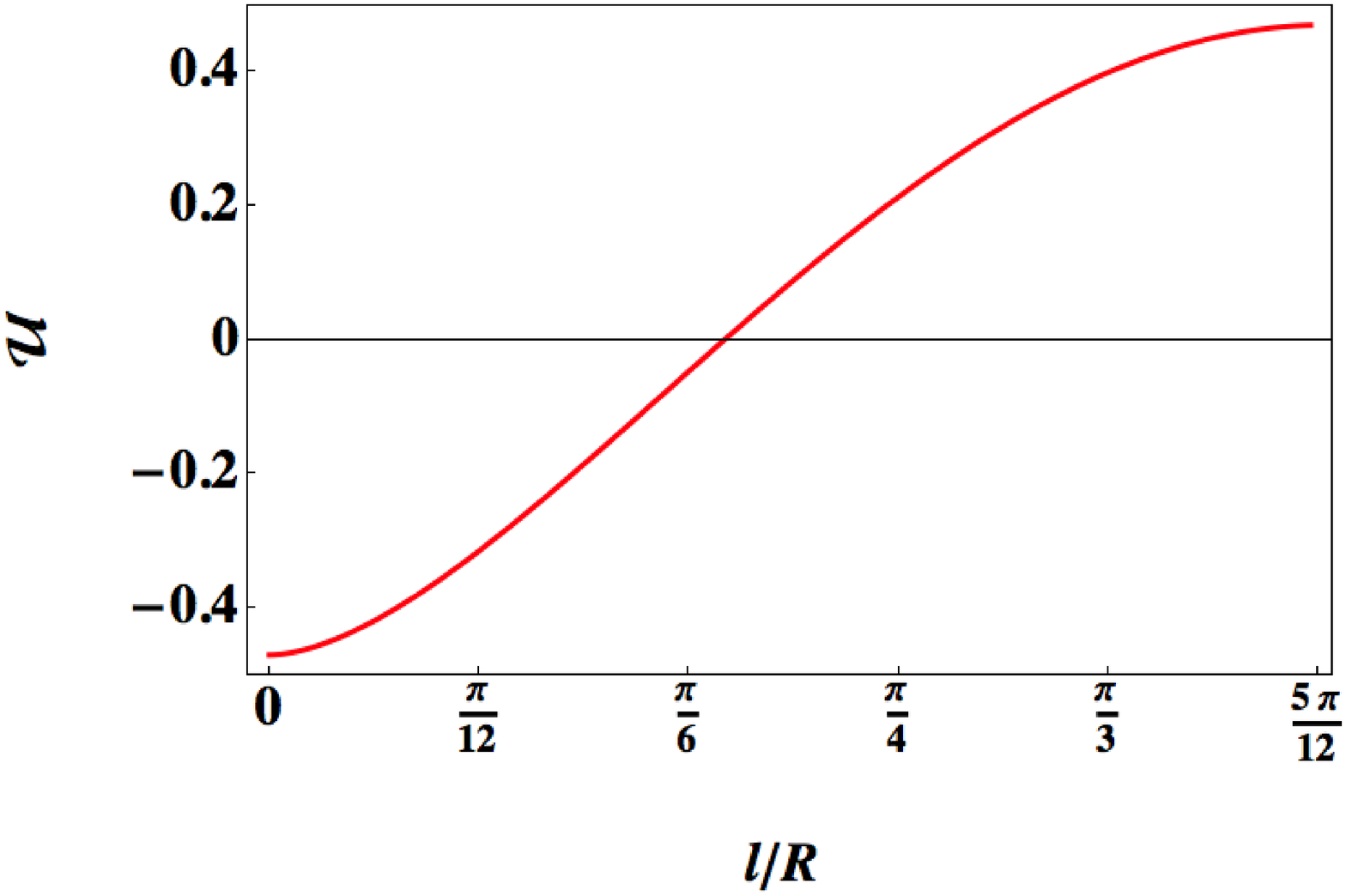} 
\end{minipage}
\begin{minipage}[t]{8cm}
\centering
\includegraphics[scale=0.28]{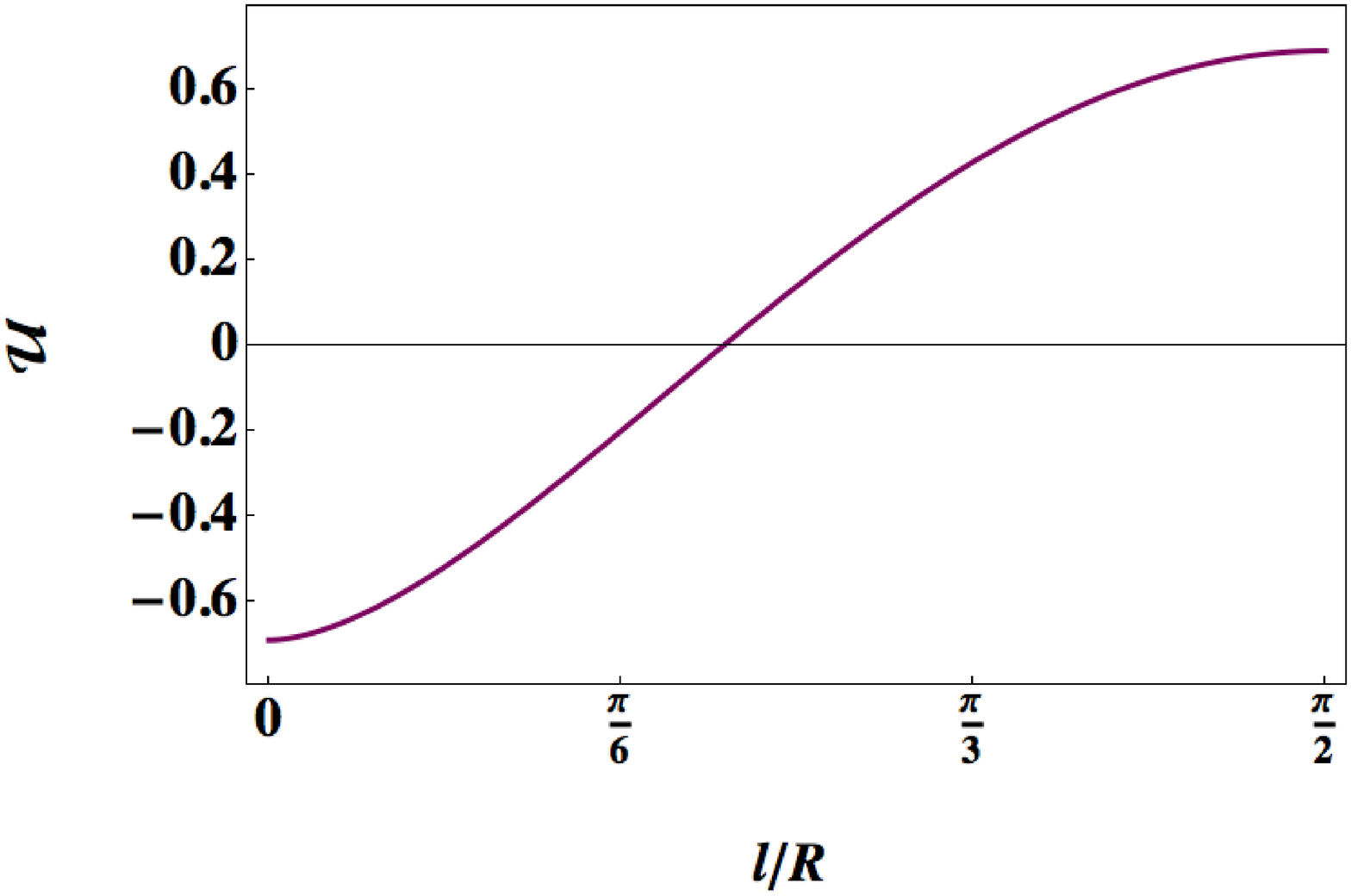}
\end{minipage}
\caption{The geometric potencial ${\cal U}$ for the spherical caps in Fig.\eqref{SPP}. The point $l=0$ being the north pole, where the defect is placed.
The force ${\cal R}_G{\cal U}$ before than the root $x_0=l_0/R $ is attractive to the defect point and repulsive after this point. }
\label{CT}
\end{center} 
\end{figure*}

\section{Summary}\label{SUMM}
In this work we have introduced a framework to calculated both  the stress tensor and  the torque induced by 
nematic ordering on curved membranes.  Using the variational principle and  differential geometry of surfaces, 
we obtain the Euler-Lagrange equations and  boundary conditions.
Taking advantage of  invariance  under translations and rotations, we find the  corresponding 
Noether charges; from these we obtain the stress tensor and the torque respectively. We find 
repulsive (attractive) forces  between defects with like (unlike) charge; defects are attracted  to points with the same sign of
gaussian curvature.  These forces are mediated by the Green function 
of the Laplace-Beltrami operator of the surface.  
Furthermore, we find  anisotropic forces that involve derivatives of  both, the Green function and  the gaussian curvature.
Extrinsic geometry only plays a role into the forces along the normal direction to the surface. 
We present these results in a coordinate independent way. 
We next applied this framework to the case of membranes with axially symmetry to analyze the spherical case. 
For a spherical vesicle with defects at the poles we find   the modified  Young-Laplace law.
We find that for certain liquid crystals, the nematic corrections to  the Young-Laplace law will be at least $50\%$, if the radius of the vesicle 
$R\sim 1-10 \mu m$,   a reasonable size in micropipette experiments.
For spherical layers with a defect at the north pole we find that the force at any point is repulsive with respect to the pole, which implies that it is an unstable equilibrium point.\\
It is possible that this {\it nematic} force be relevant in the description of nanoparticles  embedded onto spherical nematic vesicles
\cite{nelson-tetra}.
As we will show in a future report, it is possible to extend this theoretical framework to take into  account the effect of extrinsic 
couplings, a fact that may be relevant for both the texture of the nematic and  the membrane shape itself \cite{pavel}.

\section*{Acknowledgements} 

I would like to thank G. Barrientos and G. Chac\'on the interest and critical reading of the paper. I also thank Tania, 
Claus and Mario Cital\'an for having encouraged me along the way.

\smallskip
\appendix
\section{  Tangential deformation of scalar curvature}\label{APA}

We need
the deformation of the scalar curvature, 
\begin{equation}
\delta{\cal R}= g^{ab}\delta {\cal R}_{ab} +\delta g^{ab} {\cal R}_{ab}.\label{dR}
\end{equation}
The first term  in eq.(\ref{dR}) can be calculated in terms of deformations of the Christoffel symbols
\begin{equation}
g^{ab}\delta{\cal R}_{ab}=g^{ab} \nabla_c(\delta\Gamma^c_{ab}) -g^{ab} \nabla_b (\delta \Gamma^c_{ca}),\label{gdeltaR}
\end{equation}
where we can write
\begin{equation}
\delta\Gamma^c_{ab}=\frac{1}{2} g^{cd} \left( \nabla_b \delta g_{ad} +\nabla_a \delta g_{bd}-\nabla_d\delta g_{ab} \right).
\end{equation}
Using  the fact that 
the induced metric transforms as
$
\delta_{\parallel} g_{ab}=\nabla_a \Phi_b +\nabla_b \Phi_a,
$
so that  the tangential deformation of the Christoffel symbols are given by
\begin{eqnarray}
g^{ab}\delta_{\parallel}\Gamma^c_{ab}&=&\frac{1}{2} g^{cd} \left( 2\nabla^2 \Phi_d + \left[ \nabla_a, \nabla_d \right] \Phi^a 
+ \left[  \nabla_b, \nabla_d\right] \Phi^b \right)\nonumber\\
&=& \nabla^2 \Phi^c +{\cal R}_ a{}^c \Phi^a\\
g^{ab}\delta_{\parallel} \Gamma^c_{ca}&=&\frac{1}{2} g^{cd} g^{ab}  (   \left[  \nabla_c,\nabla_d  \right]  \Phi_a
+ \left[\nabla_a, \nabla_d \right] \Phi_c \nonumber\\
 &+& \left[\nabla_a, \nabla_c \right] \Phi_d
+  2 \nabla_c\nabla_a\Phi_d )\nonumber\\
&=& g^{cd}g^{ab} \nabla_c\nabla_a\Phi_d -{\cal R}_c{}^b\Phi^c,
\end{eqnarray}
where the commutator 
$ \left[ \nabla_a, \nabla_b\right] \Phi^c= {\cal R}^c{}_{dab}\Phi^d$, has been used.
By taking the corresponding gradients, 
and using the fact that $\nabla_b \nabla_c \nabla^b \Phi^c=\nabla_c \nabla^2 \Phi^c$,
we can write eq.(\ref{gdeltaR}) as
\begin{equation}
g^{ab}\delta_\parallel{\cal R}_{ab}=2\nabla_a ( {\cal R}^a_c \Phi^c).
\end{equation}
Using now this result into  (\ref{dR}) we obtain  eq.(\ref{R}).\\
\section{ The commutator $ [\delta_\parallel , \nabla^a] f $  } 
Deformation of a second derivative can be written as
\begin{equation}
[\delta_\parallel, \nabla_a \nabla_b ]f=-(\nabla_c f) \delta_\parallel \Gamma^c_{ab}, 
\end{equation}
so that
\begin{equation}
[\delta_\parallel, \nabla^2 ]f=- g^{ab}(\nabla_c f) \delta_\parallel \Gamma^c_{ab} + (\delta_\parallel g^{ab}) \nabla_a\nabla_b f, 
\end{equation}
by using $g^{ab} \delta_\parallel \Gamma^a_{bc}$  from appendix \eqref{APA},  we find Eq.\eqref{CCM}.\\
\section{Monge gauge.}
In the representation a la Monge where the embedding function is ${\bf X}(x, y)=(x, y, f(x, y)) $,   the induced metric can be written as 
$g_{ab}=\delta_{ab}+\nabla_a f\nabla_b f$ and its inverse 
\begin{equation}
g^{ab}=\delta^{ab}-\frac{ \nabla_a f\nabla_b f   }{ 1+(\nabla f)^2   }. 
\end{equation}
The normal vector to the surface 
is given by ${\bf n}=\frac{(-\nabla_a f, 1)}{\sqrt{1+(\nabla f)^2}}$. The extrinsic curvature is then
\begin{equation}
K_{ab}=-\frac{\nabla^2_{ab}f}{\sqrt{1+(\nabla f)^2}},
\end{equation}
and the mean curvature
\begin{equation}
K=-\frac{\nabla^2 f}{ \sqrt{1+(\nabla f)^2} } +\frac{  \nabla_a f\nabla_b f \nabla^2_{ab} f  } {  ( {1+(\nabla f)^2} )^{3/2}     }. 
\end{equation}
To lower order and without defects  we can write the shape equation as
\begin{eqnarray}
&&(\partial_y^2 f ) \partial_x^2 {\cal U} +  (\partial_x^2 f) \partial_y^2{\cal U} - 2 (\partial^2_{xy} f) \partial^2_{xy} {\cal U} \nonumber\\
&& + \frac{1}{2} [ \partial_x^2 f - \partial_y^2 f   ][  (\partial_y {\cal U})^2 -  (\partial_x {\cal U})^2 ]   \nonumber\\
&&- 2(\partial_{xy}^2 f) (\partial_x {\cal U}) (\partial_y {\cal U})=0.
\end{eqnarray}
When the corresponding term of the bending energy is added, the von K\'arm\'an equation is obtained.

\section{Deformation of the nematic energy}
Write the nematic energy of the membrane ${\cal \chi} $,
\begin{equation}
F= -\int_{\cal M} dA \, \chi\, \nabla^2 \chi \label{FFNN}
\end{equation}
where the field $\chi$ satisfies the equation
\begin{equation}
-\nabla^2\chi = \rho_D- {\cal R}_G, 
\end{equation}
and $\rho_D$ is the charge density. 
Deformation of \eqref{FFNN} can be written as
\begin{equation}
\delta F= - \int_{\cal M} [(\delta dA) \chi \nabla^2 \chi \underbrace{-  dA(\delta\chi) \nabla^2 \chi }_{II}\label{DDF}
\underbrace{ -  dA \chi ( \delta \nabla^2\chi)}_{III}].
\end{equation}
In the second term, deformation
of the field $\delta\chi$ can be calculated as
\begin{equation}
\delta \chi = \int_{\cal M'} dA' G(\xi, \xi') ( J' + \delta\rho_D' -\delta {\cal R}_G' ), \label{NAACHI}
\end{equation}
that is because
$
-\delta \nabla^2 \chi= \delta \rho_D - \delta {\cal R}_G, 
$	
so that if  the commutator $ [\delta, \nabla^2 ] f= J $,  we have
$
-\nabla^2\delta \chi = J+\delta \rho_D -\delta{\cal R}_G, 
$
and thus eq.\eqref{NAACHI} follows. The  integrals in eq.\eqref{DDF} can then be written as
\begin{eqnarray}
II&=&\int_{\cal M}   dA\,  \chi\,  (J +\delta\rho_D -\delta{\cal R}_G ), \nonumber\\
III&=& \int_{\cal M} dA \chi \,  ( \delta\rho_D - \delta{\cal R}_G  ). 
\end{eqnarray} 
We have then
$
II+III= \int_{\cal M} dA \, \chi (  J+ 2\delta\rho_D -2\delta {\cal R}_G ),
$
and therefore  we  can write
\begin{equation}
\delta F=  - \int_{\cal M} (\delta dA) \chi \nabla^2 \chi  +  \int_{\cal M} dA \, \chi (  J+ 2\delta\rho_D -2\delta {\cal R}_G ).\label{SSD}
\end{equation}
Once again, let us calculate separately. For the normal deformation, the first integral in eq.\eqref{SSD} becomes
\begin{equation}
-\int_{\cal M} \, (\delta_\perp dA)  \chi \nabla^2 \chi  = -\int_{\cal M} dA [K  \chi \nabla^2 \chi ] \, \Phi.  \nonumber\\
\end{equation}
In the the second integral, we substitute $J_{\perp}$ and several integrations by parts to obtain
\begin{eqnarray}
\int_{\cal M} dA\,  J_\perp \chi &=& -\int_{\cal M} dA \chi \{ [ 2K^{ab}  \nabla_a\nabla_b \chi + (\nabla_a K)(\nabla^a\chi)]  \Phi \nonumber\\
&+& (2K^{ab}- Kg^{ab}) \nabla_a\chi  \nabla_b\Phi ] \} ,\nonumber
\end{eqnarray}\\
it can be written as
\begin{eqnarray}
&=& -\int_{\cal M} dA\,  \chi [ 2K^{ab}  \nabla_a\nabla_b \chi + (\nabla_a K)(\nabla^a\chi)]  \Phi \nonumber\\
&& +\int_{\cal M} dA \nabla_b [ (2K^{ab} -Kg^{ab})\chi \nabla_a\chi] \Phi\nonumber\\
&&- \int_{\cal M} dA \nabla_b [ (2K^{ab}- Kg^{ab}) \chi (\nabla_a \chi) \Phi )    ].
\end{eqnarray}
We also have that
\begin{equation}
2\int_{\cal M} dA \, \, \chi\,  \delta_\perp \rho_D = -2\int_{\cal M} dA \chi \rho_D K \Phi. \\
\end{equation}
The last integral  in eq.\eqref{SSD} can be calculated as
\begin{eqnarray}
-2\int dA_{\cal M} \, \chi \, \delta_{\perp} {\cal R}_G &=& 2 \int_{\cal M} dA \chi\,  {\cal R}_G K\Phi \nonumber\\
&&- 2\int_{\cal M} dA\, \chi\, ( K^{ab} - g^{ab}K ) \nabla_a\nabla_b \Phi,   \nonumber
\end{eqnarray}
and after some integrations by parts we get
\begin{eqnarray}
&&= 2 \int_{\cal M} dA [ {\cal R}_G K \chi - (K^{ab}- g^{ab}K )\nabla_a\nabla_b \chi ] \Phi  \nonumber\\
&&- \int_{\cal M} dA\, 2 \, \nabla_a [(K^{ab} - g^{ab}K) ( \chi \nabla_b \Phi - \Phi\nabla_b\chi) ].
\end{eqnarray}
The normal deformation is therefore
\begin{equation}
\delta_\perp F=  \int_{\cal M} dA \, {\cal E}_\perp \Phi + \int_{\cal M} dA \nabla_a Q^a_\perp.
\end{equation}
where the normal  Euler-Lagrange derivative and the Noether charge are given respectively by
\begin{eqnarray}
{\cal E}_\perp &=&  -K  \chi \nabla^2 \chi  -   \chi [ 2K^{ab}  \nabla_a\nabla_b \chi + (\nabla_a K)(\nabla^a\chi)] \nonumber\\
&&+\nabla_b [ (2K^{ab} -Kg^{ab})\chi \nabla_a\chi] - 2 \chi \rho_D K   \nonumber\\
&&+ 2[ {\cal R}_G K \chi - (K^{ab}- g^{ab}K )\nabla_a\nabla_b \chi  ] \nonumber\\
&=& 2( K g^{ab} - K^{ab})  \nabla_a\nabla_b \chi + ( 2K^{ab} -Kg^{ab}  ) \nabla_a\chi \nabla_b \chi, \nonumber\\
Q^a_\perp &=&-(2K^{ab}- Kg^{ab}) \chi (\nabla_a \chi) \Phi \nonumber\\
&& -2(K^{ab} - g^{ab}K) ( \chi \nabla_b \Phi - \Phi\nabla_b\chi).
\end{eqnarray}
The tangential  deformation can be calculated in a similar way. By using the tangential deformation
of the area we have
\begin{eqnarray}
&&-\int_{\cal M} \, (\delta_\parallel dA)  \chi \nabla^2 \chi  = -\int_{\cal M} dA \,\chi \, \nabla^2 \chi ( \nabla_a \Phi^a)\nonumber \\
&&= -\int_{\cal M} dA \nabla_a ( \chi \nabla^2 \chi \Phi^a ) + \int_{\cal M} dA \nabla_a( \chi \nabla^2 \chi ) \Phi^a. 
\end{eqnarray}
We also obtain that the integral
\begin{eqnarray}
\int_{\cal M} dA\,  J_\parallel \chi &=& \int_{\cal M} dA \chi [ ( -\nabla^2 \Phi^a + {\cal R}_G \Phi^a )\nabla_a {\chi} \nonumber\\
&&- 2 (\nabla^a\Phi^b )\nabla_a\nabla_b {\chi}   ], \nonumber 
\end{eqnarray}
can be rewritten after integrations by parts
\begin{eqnarray}
&&=-\int dA_{\cal M} \nabla^a [ \nabla_a\Phi^b  \chi \nabla_b\chi ] + \int_{\cal M} dA\nabla_a [ \Phi^b \nabla^a(\chi \nabla_b\chi  )  ]  \nonumber\\
&&- \int_{\cal M} dA \Phi^a \nabla^2 ( \chi \nabla_a \chi ) +\int_{\cal M} dA \Phi^a {\cal R}_G \chi\nabla_a\chi \nonumber\\
&&- 2\int_{\cal M} dA \nabla^a( \Phi^b\chi \nabla_a\nabla_b \chi  )  + 2\int_{\cal M} dA \Phi^b \nabla^a ( \chi \nabla_a\nabla_b \chi  ).   \nonumber\\
\end{eqnarray}
The next integration can be done as
\begin{eqnarray}
&&2\int_{\cal M} dA \, \, \chi\,  \delta_\parallel \rho_D = -2\int_{\cal M} dA\, \chi \, \rho_D \nabla_a\Phi^a \nonumber\\
&&=-2 \int_{\cal M} dA \nabla_a (\chi \rho_D \Phi^a ) +2\int_{\cal M} dA \nabla_a (\chi \rho_D ) \Phi^a\nonumber \\
\end{eqnarray}
and finally we get
\begin{equation}
-2\int_{\cal M} dA \, \chi \, \delta_{\parallel} {\cal R}_G =-2 \int_{\cal M} dA\, \chi \, \Phi^a \nabla_a {\cal R}_G. 
\end{equation}
So that we obtain the tangential derivative and the Noether charge as
\begin{eqnarray}
{\cal E}_a &=& \nabla_a( \chi \nabla^2 \chi ) - \nabla^2 ( \chi \nabla_a \chi ) +{\cal R}_G \chi\nabla_a\chi \nonumber\\
&&	 + 2 \nabla^b ( \chi \nabla_b\nabla_a \chi  ) +2 \nabla_a (\chi \rho_D ) -2  \chi \,  \nabla_a {\cal R}_G \nonumber\\
&=& 2(\rho_D + {\cal R}_G \chi ) \nabla_a \chi.\nonumber\\
Q_\parallel^a&=& -( \chi \nabla^2 \chi \, \Phi^a ) -[ \nabla^a\Phi^b \, \chi \nabla_b\chi ] +  [ \Phi^b \nabla^a (\chi \nabla_b \chi ) ] \nonumber\\
&&-2( \Phi^b \chi \nabla_a\nabla_b \chi )
-2 ( \chi\,  \rho_D \,  \Phi^a ).\nonumber\\
&=& \Phi^b[ \nabla^a(\chi \nabla_b\chi ) - 2\chi \nabla^a\nabla_b\chi - \delta^a_b (\rho_D + {\cal R}_G ) \chi   ] \nonumber\\
&&-\chi \nabla_b\chi \nabla^a\Phi^b.
\end{eqnarray}

\section{Darboux frame}\label{DARFRA}
For the second integral we recall the Darboux basis adapted to the boundary ${\cal C}$ parametrized by arc length.
Define ${\bf T}$  its tangent vector such that ${\bf T}= T^a{\bf e}_a$,
we also define ${\bf l}= {\bf T}\times {\bf n} $ the normal unit to the boundary, tangent
to the surface. We have that
\begin{eqnarray}
\dot{\bf T} &=& \kappa_n {\bf n} + \kappa_g {\bf l}, \nonumber\\
\dot{\bf l} &=& -\kappa_g {\bf T} - \tau_g {\bf n}, \nonumber\\
\dot{\bf n} &=& -\kappa_n {\bf T}+\tau_g {\bf l}. \label{DAR}
\end{eqnarray}
In these equations, we have defined the normal curvature 
\begin{eqnarray}
\kappa_n &=&\dot{\bf T}\cdot {\bf n}, \nonumber\\
&=&( \dot T^a {\bf e}_a - K_{ab}T^a  T^b {\bf n} )\cdot {\bf n },\nonumber\\
&=&-K_{ab}T^a T^b,
\end{eqnarray}
and its geodesic curvature
\begin{eqnarray}
\kappa_g &=&\dot{\bf T}\cdot {\bf l}, \nonumber\\
&=& \kappa_g^a {\bf e}_a\cdot {\bf l},\nonumber\\
&=&(\dot T^a + \Gamma^a_{bc}T^bT^c )l_a.
\end{eqnarray}
The second equation in \eqref{DAR} defines the geodesic torsion
\begin{eqnarray}
\tau_g &=&\dot{\bf n} \cdot {\bf l}, \nonumber\\
&=& K_{ab} T^al^b.
\end{eqnarray}
Let us  calculate the deformations in the Darboux frame.
Deformation of the boundary  is given by 
\begin{eqnarray}
\delta{ \bf X} &=& \phi {\bf T}+ \psi {\bf l} + \Phi {\bf n} ,\nonumber\\
&=& \Phi^a {\bf e}_a + \Phi {\bf n}.
\end{eqnarray}
that is $\Phi^a T_a = \phi$ and $\Phi^a l_a=\psi$.
Therefore, deformation of the unit tangent can be written as
\begin{eqnarray}
\delta {\bf T} &=&\dot\phi {\bf T} + \dot{\psi} {\bf l} + \dot\Phi {\bf n}
+ \phi \dot{\bf T} + \psi \dot{\bf l} + \Phi \dot{\bf n}, \nonumber\\
&=& ( \dot\phi - \kappa_g\psi -\kappa_n \Phi   ) {\bf T} + ( \dot\psi +\kappa_g \phi + \tau_g \Phi    ){\bf l} \nonumber\\
&+& (\dot\Phi + 	\kappa_n\phi - \tau_g \psi ) {\bf n}.
\end{eqnarray}
Then we obtain
\begin{eqnarray}
\delta\oint_{\cal C}\, ds &=& \oint_{\cal C}\, ds\,  {\bf T} \cdot\delta {\bf T }, \nonumber\\
&=& \oint_{\cal C} ds\, ( \dot\phi - \kappa_g\psi -\kappa_n \Phi   ), \nonumber\\
\delta L&=& \Delta\phi- \oint_{\cal C} ds\, (  \kappa_g\psi + \kappa_n \Phi   ).
\end{eqnarray}
where $\Delta\phi= 0$ for a closed curve.  Thus, $\delta L $
does not include deformation along the unit tangential  vector.
Write 
\begin{eqnarray}
Q_\perp ^a&=& M^{ab} \nabla_b\Phi + M^a \Phi\nonumber\\
Q_\parallel^a&=& N^a{}_b\Phi^b + N_b\nabla^a\Phi^b,
\end{eqnarray}
where 
\begin{eqnarray}
M^{ab}&=&2 ( Kg^{ab}- K^{ab} ) \chi \nonumber\\
M^a&=&[( Kg^{ab} - K^{ab} ) (\chi -2) - K^{ab}\chi  ] \nabla_b \chi, \nonumber\\
N^{ab}&=&  \nabla^a\chi\nabla^b\chi - \chi \nabla^a\nabla^b\chi - g^{ab}(\rho_D+ {\cal R}_G )\chi, \nonumber\\
N^a&=& -\chi \nabla^a \chi. 
\end{eqnarray}
The we can obtain
\begin{eqnarray}
l_a Q^a &=&l_a(Q_\perp^a + Q_\parallel^a  ) \nonumber\\
&=&l_a ( M^{ab}\nabla_b  \Phi + M^a\Phi )  + l_a ( N^{ab} \Phi_b + N^b \nabla^a \Phi_b) , \nonumber\\
&=&l_aM^{ab}l_b \nabla_l\Phi + l_a M^{ab}T_b \dot\Phi + l_a M^a\Phi\nonumber\\
&+&( l_a N^{ab}T_b  + N^b \nabla_l T_b       ) \,  \phi + ( l_a N^{ab}l_b + N^b \nabla_l l_b    )\psi \nonumber\\
&+& N^b l_b \nabla_l \psi + N^bT_b \nabla_l \phi, 
\end{eqnarray}
where we have used that on the boundary 
\begin{eqnarray}
\nabla_a\Phi &=& {\bf e}_a \cdot \nabla \Phi, \nonumber\\
&=& ( l_a{\bf l} + T_a {\bf T}  ) \cdot \nabla\Phi, \nonumber\\
&=&l_a \nabla_l \Phi + T_a \dot\Phi, 
\end{eqnarray}
that is $\nabla_l \Phi= l^a\nabla_a \Phi $ and $\dot\Phi= T^a \nabla_a \Phi $.
We also have that
\begin{equation}
\nabla_b\Phi^a=T_b \dot\Phi^a + l_b\nabla_l \Phi^a.
\end{equation}
Note us that on the boundary, the independent  deformations are given by the scalars functions $ \psi , \phi, \Phi $.
Then we have that
\begin{equation}
\delta{\cal H}= \oint ds [ \kappa_Al_a Q^a + (\sigma -\sigma_b \kappa_g )\psi -\sigma_b \kappa_n \Phi ].
\end{equation}

\section{Green function and geometric potential on the sphere.}\label{GREEN}
In order to find the Green function on the sphere, we need
\begin{equation}
I=\int_0^{\pi R} d\ell \rho (\ell) \log r_>.
\end{equation}
We split the integral as 
\begin{eqnarray}
I &=&\log r(l) \int_0^l d\ell \rho(\ell) + \int_l^{\pi R}d\ell \,\, \rho(\ell) \log r(\ell ),\nonumber\\
&=&\log [R\tan(l/2R)] \int_0^{l} d\ell R\sin ( \ell /R) \nonumber\\
&&+ \int _l^{\pi R} d\ell \,  R\sin ( \ell/R) \log[ R\tan (\ell /2R)].\nonumber\\
&=&-R^2\log[ R\tan (l/2R)] [ \cos(l/R)-1 ] \nonumber\\
&&+ R^2  \int_{l/R}^\pi dx \sin x \log( R\tan x/2 ). 
\end{eqnarray}
Here, the integral can be obtained as
\begin{eqnarray} 
&&\int_{l/R}^\pi dx \sin x \log( R\tan x/2 ) = \log R \nonumber\\
&-& \log[ \sin(l/2R)\cos(l/2R)] \nonumber\\
&+& \cos(l/R) \log[ R\tan (l/2R)].
\end{eqnarray}
When substituting we obtain \eqref{GAB}.
The Green function is then given by
\begin{eqnarray}
G(\xi, \zeta) &=& -\frac{1}{4\pi} \log [r(l)^2 +r(\ell)^2 -2r(l)r(\ell) \cos (\phi-\varphi ) ] \nonumber\\
&&-\frac{1}{4\pi} \log[  \cos^2(l/2R)\cos^2(\ell /2R)  ], \label{GRR}
\end{eqnarray}
that no longer contains singularities.
By using the Green function eq.\eqref{GRR}, we can evaluate the geometric potential as
\begin{equation}
{\cal U}(\xi)= \int dA_\zeta G(\xi , \zeta) R_G(\ell).
\end{equation}
The gaussian curvature of the sphere is given by ${\cal R}_G=1/R^2$, such that
\begin{equation}
{\cal U}=\frac{1}{R^2}  \int_0^{\pi R} d\ell \rho(\ell) \int_0^{2\pi} d\varphi \, \, G(\xi, \zeta ). 
\end{equation}
As an intermediate step we obtain
\begin{equation}
{\cal U}= \log\cos^2 (l/2R) - I_1, \label{UU}
\end{equation}
where $I_1$ is written as
\begin{equation}
I_1=\frac{1}{4\pi R^2} \int dA_\zeta \log[ \cos^2 (l/2R) \cos^2( \ell/2R)].
\end{equation}
We split this integral as
\begin{eqnarray}
4\pi R^2 I_1&=& 2\pi \log \cos^2 (l/2R)\int_0^{\pi R} d\ell \rho (\ell) \nonumber\\
&&+ 2\pi\int_0^{\pi R} d\ell \rho (\ell)  \log \cos^2 (\ell /2R)   \nonumber\\
&=& 4\pi R^2 \log \cos^2 (l/2R) -4\pi R^2.
\end{eqnarray}
in such a way that when substituting into \eqref{UU} we get  ${\cal U}=1. $

\end{document}